\documentclass[twocolumn]{aastex63}

\newcounter{daggerfootnote}

\newcommand{\oiii}{O\,\textsc{iii}}
\newcommand{\oii}{O\,\textsc{ii}}

\newcommand{\nev}{[Ne\,\textsc{v}]}
\newcommand{\heii}{He\,\textsc{ii}}

\newcommand{\muv}{$M_{\mathrm {UV}}$}
\newcommand{\jwst}{\textit{JWST}}
\newcommand{\hst}{\textit{HST}}

\definecolor{mycol}{rgb}{0,0,1}

\received{n/a}
\revised{n/a}
\accepted{n/a}

\submitjournal{\apjl}

\shorttitle{CAPERS High-$z$}
\shortauthors{Kokorev et al.}

\usepackage{xcolor}
\usepackage{longtable}
\usepackage{lipsum}
\usepackage{amsmath}
\usepackage[normalem]{ulem}
\usepackage[para]{threeparttable}
\usepackage{enumitem}
\usepackage[encapsulated]{CJK}

\begin{document}

\title{CAPERS Observations of Two UV-Bright Galaxies at \boldmath{$z>10$}. \\ More Evidence for Bursting Star Formation in the Early Universe.}

\correspondingauthor{Vasily Kokorev}
\email{vkokorev@utexas.edu}

\author[0000-0002-5588-9156]{Vasily Kokorev}
\affiliation{Department of Astronomy, The University of Texas at Austin, Austin, TX 78712, USA}

\author[0000-0003-2332-5505]{\'Oscar A. Ch\'avez Ortiz}
\affiliation{Department of Astronomy, The University of Texas at Austin, Austin, TX 78712, USA}

\author[0000-0003-1282-7454]{Anthony J. Taylor}
\affiliation{Department of Astronomy, The University of Texas at Austin, Austin, TX 78712, USA}

\author[0000-0001-8519-1130]{Steven L. Finkelstein}
\affiliation{Department of Astronomy, The University of Texas at Austin, Austin, TX 78712, USA}

\author[0000-0002-7959-8783]{Pablo Arrabal Haro}
\altaffiliation{NASA Postdoctoral Fellow}
\affiliation{Astrophysics Science Division, NASA Goddard Space Flight Center, 8800 Greenbelt Rd, Greenbelt, MD 20771, USA}

\author[0000-0001-5414-5131]{Mark Dickinson}
\affiliation{NSF's National Optical-Infrared Astronomy Research Laboratory, 950 N. Cherry Ave., Tucson, AZ 85719, USA}

\author[0000-0002-0302-2577]{John Chisholm}
\affiliation{Department of Astronomy, The University of Texas at Austin, Austin, TX 78712, USA}

\author[0000-0001-7201-5066]{Seiji Fujimoto}
\affiliation{Department of Astronomy, The University of Texas at Austin, Austin, TX 78712, USA}

\author[0000-0002-8984-0465]{Julian B.~Mu\~noz}
\affiliation{Department of Astronomy, The University of Texas at Austin, Austin, TX 78712, USA}

\author[0000-0003-4564-2771]{Ryan Endsley}
\affiliation{Department of Astronomy, The University of Texas at Austin, Austin, TX 78712, USA}

\author[0000-0003-3424-3230]{Weida Hu}
\affiliation{Department of Physics and Astronomy, Texas A\&M University, College Station, TX 77843-4242, USA}
\affiliation{George P. and Cynthia Woods Mitchell Institute for Fundamental Physics and Astronomy, Texas A\&M University, College Station, TX 77843-4242, USA}

\author[0000-0002-8951-4408]{Lorenzo Napolitano}
\affiliation{INAF – Osservatorio Astronomico di Roma, via Frascati 33, 00078, Monteporzio Catone, Italy}
\affiliation{Dipartimento di Fisica, Università di Roma Sapienza, Città Universitaria di Roma - Sapienza, Piazzale Aldo Moro, 2, 00185, Roma,
Italy}

\author[0000-0003-3903-6935]{Stephen M.~Wilkins} 
\affiliation{Astronomy Centre, University of Sussex, Falmer, Brighton BN1 9QH, UK}
\affiliation{Institute of Space Sciences and Astronomy, University of Malta, Msida MSD 2080, Malta}


\author[0000-0003-3596-8794]{Hollis B. Akins}
\affiliation{Department of Astronomy, The University of Texas at Austin, Austin, TX 78712, USA}

\author[0000-0001-5758-1000]{Ricardo Amori\'in}
\affiliation{Instituto de Astrof\'isica de Andaluc\'ia (CSIC), Apartado 3004, 18080 Granada, Spain}

\author[0000-0002-0930-6466]{Caitlin M. Casey}
\affiliation{Department of Physics, University of California, Santa Barbara, CA 93106, USA}
\affiliation{Cosmic Dawn Center (DAWN), Niels Bohr Institute, University of Copenhagen, Jagtvej 128, K{\o}benhavn N, DK-2200, Denmark}

\author[0000-0001-8551-071X]{Yingjie Cheng}
\affiliation{University of Massachusetts Amherst, 710 North Pleasant Street, Amherst, MA 01003-9305, USA}

\author[0000-0001-7151-009X]{Nikko J. Cleri}
\affiliation{Department of Astronomy and Astrophysics, The Pennsylvania State University, University Park, PA 16802, USA}
\affiliation{Institute for Computational and Data Sciences, The Pennsylvania State University, University Park, PA 16802, USA}
\affiliation{Institute for Gravitation and the Cosmos, The Pennsylvania State University, University Park, PA 16802, USA}

\author[0000-0002-6348-1900]{Justin Cole}
\affiliation{Department of Physics and Astronomy, Texas A\&M University, College Station, TX 77843-4242, USA}
\affiliation{George P. and Cynthia Woods Mitchell Institute for Fundamental Physics and Astronomy, Texas A\&M University, College Station, TX 77843-4242, USA}

\author[0000-0002-3736-476X]{Fergus Cullen}
\affiliation{Institute for Astronomy, University of Edinburgh, Royal Observatory, Edinburgh EH9 3HJ, UK}

\author[0000-0002-3331-9590]{Emanuele Daddi}
\affiliation{Universit\'e Paris-Saclay, Universit\'e Paris Cit\'e, CEA, CNRS, AIM, 91191 Gif-sur-Yvette, France}

\author[0000-0001-8047-8351]{Kelcey Davis}
\affiliation{Department of Physics, 196A Auditorium Road, Unit 3046, University of Connecticut, Storrs, CT 06269, USA}

\author[0000-0002-7622-0208]{Callum T. Donnan}
\affiliation{NSF's National Optical-Infrared Astronomy Research Laboratory, 950 N. Cherry Ave., Tucson, AZ 85719, USA}

\author[0000-0002-1404-5950]{James S. Dunlop}
\affiliation{Institute for Astronomy, University of Edinburgh, Royal Observatory, Edinburgh EH9 3HJ, UK}

\author[0000-0003-0531-5450]{Vital Fern\'andez}
\affiliation{Michigan Institute for Data Science, University of Michigan, 500 Church Street, Ann Arbor, MI 48109, USA}

\author[0000-0002-7831-8751]{Mauro Giavalisco}
\affiliation{University of Massachusetts Amherst, 710 North Pleasant Street, Amherst, MA 01003-9305, USA}

\author[0000-0001-9440-8872]{Norman A. Grogin}
\affiliation{Space Telescope Science Institute, 3700 San Martin Drive, Baltimore, MD 21218, USA}

\author[0000-0001-6145-5090]{Nimish Hathi}
\affiliation{Space Telescope Science Institute, 3700 San Martin Drive, Baltimore, MD 21218, USA}

\author[0000-0002-3301-3321]{Michaela Hirschmann}
\affiliation{Institute of Physics, Laboratory for Galaxy Evolution, Ecole Polytechnique Federale de Lausanne, Observatoire de Sauverny, Chemin Pegasi 51, 1290 Versoix, Switzerland}

\author[0000-0001-9187-3605]{Jeyhan S. Kartaltepe}
\affiliation{Laboratory for Multiwavelength Astrophysics, School of Physics and Astronomy, Rochester Institute of Technology, 84 Lomb Memorial Drive, Rochester, NY
14623, USA}

\author[0000-0002-6610-2048]{Anton M. Koekemoer}
\affiliation{Space Telescope Science Institute, 3700 San Martin Drive, Baltimore, MD 21218, USA}

\author[0000-0003-0486-5178]{Ho-Hin Leung}
\affiliation{Institute for Astronomy, University of Edinburgh, Royal Observatory, Edinburgh EH9 3HJ, UK}

\author[0000-0003-1581-7825]{Ray A. Lucas}
\affiliation{Space Telescope Science Institute, 3700 San Martin Drive, Baltimore, MD 21218, USA}

\author[0000-0003-4368-3326]{Derek McLeod}
\affiliation{Institute for Astronomy, University of Edinburgh, Royal Observatory, Edinburgh EH9 3HJ, UK}

\author[0000-0001-7503-8482]{Casey Papovich}
\affiliation{Department of Physics and Astronomy, Texas A\&M University, College
Station, TX, 77843-4242 USA}
\affiliation{George P.\ and Cynthia Woods Mitchell Institute for
 Fundamental Physics and Astronomy, Texas A\&M University, College
 Station, TX, 77843-4242 USA}
 
\author[0000-0001-8940-6768]{Laura Pentericci}
\affiliation{INAF – Osservatorio Astronomico di Roma, via Frascati 33, 00078, Monteporzio Catone, Italy}

\author[0000-0003-4528-5639]{Pablo G. P\'erez-Gonz\'alez}
\affiliation{Centro de Astrobiolog\'{\i}a (CAB), CSIC-INTA, Ctra. de Ajalvir km 4, Torrej\'on de Ardoz, E-28850, Madrid, Spain}

\author[0000-0002-6748-6821]{Rachel S. Somerville}
\affiliation{Center for Computational Astrophysics, Flatiron Institute, 162 5th Avenue, New York, NY, 10010, USA}

\author[0000-0002-9373-3865]{Xin Wang}
\affiliation{School of Astronomy and Space Science, University of Chinese Academy of Sciences (UCAS), Beijing 100049, China}
\affiliation{National Astronomical Observatories, Chinese Academy of Sciences, Beijing 100101, China}
\affiliation{Institute for Frontiers in Astronomy and Astrophysics, Beijing Normal University, Beijing 102206, China}

\author[0000-0003-3466-035X]{{L. Y. Aaron} {Yung}}
\affiliation{Space Telescope Science Institute, 3700 San Martin Drive, Baltimore, MD 21218, USA}

\author[0000-0002-7051-1100]{Jorge A. Zavala}
\affiliation{University of Massachusetts Amherst, 
710 North Pleasant Street, Amherst, MA 01003-9305, USA}

\begin{abstract}
We present the first results from the CAPERS survey, utilizing PRISM observations with the \textit{JWST}/NIRSpec MSA in the PRIMER-UDS field. With just 14\% of the total planned data volume, we spectroscopically confirm two new bright galaxies ($M_{\rm UV}\sim -20.4$) at redshifts $z = 10.562\pm0.034$ and $z = 11.013\pm0.028$. We examine their physical properties, morphologies, and star formation histories, finding evidence for recent bursting star formation in at least one galaxy thanks to the detection of strong (EW$_0\sim70$ \AA) H$\gamma$ emission.
Combining our findings with previous studies of similarly bright objects at high-$z$, we further assess the role of stochastic star formation processes in shaping early galaxy populations. Our analysis finds that the majority of bright ($M_{\rm UV}\lesssim -20$) spectroscopically-confirmed galaxies at $z>10$ were likely observed during a starburst episode, characterized by a median SFR$_{10}$/SFR$_{100}\sim2$, although with substantial scatter. Our work also finds tentative evidence that $z>10$ galaxies are more preferentially in a bursting phase than similarly bright $z\sim6$ galaxies. We finally discuss the prospects of deeper spectroscopic observations of a statistically significant number of bright galaxies to quantify the true impact of bursting star formation on the evolution of the bright end of the ultraviolet luminosity function at these early epochs. 
\end{abstract}


\section{Introduction} \label{sec:intro}
Probing the earliest stages of galaxy formation and evolution, within the first few hundred million years of cosmic history, remains one of the greatest challenges in astronomy. These primordial galaxies have long evaded direct detection due to their intrinsic faintness and extreme redshifts. Prior to the launch of the \textit{James Webb Space Telescope} (\textit{JWST}), state of the art space-based facilities such as the \textit{Hubble Space Telescope} (\textit{HST}) and \textit{Spitzer Space Telescope} (\textit{Spitzer}), only managed to capture early star formation for moderate-sized samples up to $z \sim$ 7--8 \citep[e.g.,][]{ellis13,mclure13,bouwens15,finkelstein15}, and very small samples to $\sim$ 11 \citep[e.g.,][]{oesch16,bouwens21,finkelstein22b}. Detailed spectroscopic studies of rest optical/rest UV emission at these redshifts, and any further investigations at higher redshifts remained inaccessible.

Recent \textit{JWST} observations have managed to successfully spectroscopically confirm galaxies in the early Universe, as early as $\sim280$ Myr ($z=14$) after the Big Bang \citep{arrabal_haro23,bunker23,curtislake23,carniani24,castellano24,hsiao23,napolitano25,zavala25}. While these discoveries provide an unprecedented glimpse into the emergence of the first cosmic structures, they also intensify the tension between simulations and observations. The bright (\muv $\lesssim-20$) galaxies identified in these studies appear to be unexpectedly abundant, as reflected in their contribution to the rest-frame ultraviolet luminosity function (UVLF; e.g., \citealt{finkelstein24,harikane24}), in stark contrast to most cosmological models calibrated before \textit{JWST} (e.g., see \citealt{boylan-kolchin23,dayal14,mason15,behroozi15,yung19,yung20}, although c.f. \citealt{lovell21,wilkins23}). This, in turn, suggests that the cosmic star formation rate (SFR) density at $z>10$ does not exhibit the rapid decline predicted by constant star formation efficiency models \citep{bouwens23a,bouwens23b,harikane23,oesch18}, but instead follows a shallower trajectory, consistent with earlier hints from \textit{HST} data \citep{mcleod16} and supported by recent \textit{JWST} photometric samples \citep{donnan23,donnan24,mcleod24,2023ApJ...951L...1P}.

Several explanations have been proposed for the observed over-abundance of bright galaxies at high redshifts \citep[for detailed discussions, see e.g.,][]{finkelstein24,harikane23,harikane24,casey24,franco24}. Current observations do not yet yield conclusive results on the contributions of active galactic nuclei (AGN) from galaxies at these early epochs. AGN emission, particularly from unobscured quasars, can significantly enhance observed \muv, potentially inflating the estimated abundance of bright galaxies in the UVLF. Indeed, some $M_{\text{UV}}\sim-20$ sources identified at $z>9-10$ exhibit compact morphologies accompanied by high-ionization UV spectral lines, strongly suggesting AGN activity \citep{bunker23,maiolino23,napolitano24}. However, an approximately equal number of similarly bright galaxies are spatially resolved by \textit{JWST}, displaying weak or absent AGN spectral signatures \citep{carniani24,arrabal_haro23,arrabal_haro23b}. While AGN activity may thus account for some of these bright galaxies, this alone cannot fully explain their overall abundance. 

The enhanced brightness of $z>10$ galaxies could arise from vigorous, star-formation-driven outflows expelling dust and gas from these systems, resulting in extremely blue UV colors, nearly dust-free stellar populations, and elevated UV luminosities \citep{ferrara23,ferrara25}. Most of the currently-confirmed bright high-$z$ galaxies \citep{arrabal_haro23,castellano24,hsiao23} indeed exhibit very blue UV slopes ($\beta\sim-2.7$) and minimal dust attenuation ($A_{\rm V}\sim0$), although with few notable exceptions \citep{bunker23,carniani24}. Still, the models outlined in \citet{ferrara23,ferrara25} require a rapid change of the dust attenuation (and thus UV colors) at \muv$= -20$ between $z\sim8$ and $11$; however, the observational evidence for this is still unclear and is primarily based on photometric samples \citep{finkelstein23,papovich23,cullen24}. 

It is also possible that a shift toward a more heavy initial mass function (IMF) can be present in these early galaxies, driven by both different chemical composition and the temperature of the cosmic microwave background (CMB) at high redshift. \citep{larson98,bromm02,tumlinson06,inayoshi22,steinhardt23,yung24}. Current high-$z$ observations still cannot reliably constrain the IMF, however, as this would require deep rest-UV spectra in order to detect the signatures of massive stars such as the \heii\, and \nev\, lines \citep[e.g. see][]{cleri23,chisholm24,katz23,olivier22,trussler23}.

Moreover, significantly reduced gas depletion timescales in early-Universe halos, facilitated by high gas densities and diminished effectiveness of feedback processes, could further enhance star formation efficiencies as indicated by the Feedback-Free Starburst model \citep{dekel23,li24_ffb}. 

Future studies combining data from surveys that cover very large volumes (e.g. COSMOS-Web, COSMOS-3D) will soon offer improved constraints on this topic. If indeed these observations will be able to confirm that the bright end of the UVLF flattens out, potential drivers of this evolution could be processes that preferentially impact more massive halos, as pointed out by \citet{finkelstein23}. Nevertheless, significant variations in star formation efficiency (SFE) or the IMF over the relatively limited dynamic range of \muv\, and therefore halo masses, appear unlikely \citep[e.g. see][]{donnan25}. 

Finally, the increase in stochasticity in the star formation histories (SFHs) of high-$z$ galaxies could be a potential driver behind the observed trends \citep{sun_g23,shen23,munoz23,gelli24,yung24,kravtsov24}. Introducing some dispersion into the ratio of halo mass to UV luminosity will "upscatter"
some of the numerous low-mass objects to populate the bright end of the UVLF, an effect similar to the Eddington bias. Emerging observational evidence for increased SFR variability at high-$z$ includes spectroscopic studies of lower-mass galaxies \citep{looser23}, early (likely temporarily) quiescent galaxies at $z\sim7$ \citep{looser24,weibel24}, and photometric analyses indicating recent SFH upturns in brighter galaxies at $z\sim6$ \citep{ciesla24,endlsey24}. Despite this evidence, the current scarcity of deep \textit{JWST}/PRISM spectroscopy—and consequently, limited emission line detections—hampers statistical analyses of SFHs in large samples of high-$z$ galaxies. 

In this work, we present initial NIRSpec spectroscopic observations of very high redshift galaxies from the CANDELS-Area Prism Epoch of Reionization Survey (CAPERS) program (PID \#6368, PI: M. Dickinson). 
With just the first $\sim 14$\% of the dataset collected, CAPERS has already identified two  bright ($M_{\rm UV}\sim-20.4$) galaxies at $z>10$, increasing the known population of such objects to 12. Here we discuss the properties and SFHs of these newly discovered galaxies and combine our observations with existing studies to assess the relevance of stochastic star formation in shaping galaxy populations in the early Universe.

This paper is organized as follows. In \autoref{sec:obs_data} we present the CAPERS survey alongside the \jwst\, observational data sets used in this study. In \autoref{sec:data_analysis} we calculate the spectroscopic redshifts, select $z>10$ galaxies, and discuss SED fitting and emission line measurements. In \autoref{sec:res} we discuss how our results compare to the other bright galaxies spectroscopically confirmed at high-$z$ and discuss the potential impact of bursting star-formation on the enhancement of the UVLF. We summarize our findings in \autoref{sec:concl}. Throughout this work we assume a flat $\Lambda$CDM cosmology with $\Omega_{\mathrm{m},0}=0.3$, $\Omega_{\mathrm{\Lambda},0}=0.7$ and H$_0=70$ km s$^{-1}$ Mpc$^{-1}$, and a \citet{chabrier} initial mass function (IMF) between $0.1-100$ $M_{\odot}$. All magnitudes are expressed in the AB system \citep{oke74}.

\section{Observations and Data} \label{sec:obs_data}
Detailed descriptions of CAPERS target selection and prioritization, observations, data reduction, redshift measurement and spectral analysis will be presented in future publications, and are briefly summarized below.

\subsection{CAPERS}
CAPERS is an ongoing \jwst\ Cycle 4 Treasury program of moderately deep NIRSpec multi-object spectroscopy in three fields from CANDELS \citep{grogin11,koekemoer11} with wide-area public \jwst\ NIRCam imaging:  the EGS \citep[from CEERS, PI: S.\ Finkelstein,][]{finkelstein25}, and COSMOS and UDS  (from PRIMER, PI: J.Dunlop).  CAPERS is obtaining low-resolution NIRSpec PRISM spectra for thousands of distant galaxies, including more than 100 candidates for objects at $z > 10$.  CAPERS is building a spectroscopic legacy dataset that serves as a key resource for understanding key properties of the earliest galaxies in the Universe. In our work we focus on the first batch of CAPERS observations executed in the PRIMER-UDS field.

\subsection{NIRSpec Observations}
\label{sec:nirspec_setup}
CAPERS is obtaining NIRSpec multi-object spectroscopy at seven pointing positions in each of its three survey fields. At each pointing, CAPERS normally observes three configurations of the NIRSpec Micro-Shutter Assembly (MSA) with a common pointing center and position angle, each with an effective exposure time of 5690\,s (1.58\,h). This scheme allows for faint and high-priority targets to be repeated in multiple configurations and thus receive total exposure times up to 17069\,s (4.74\,h).  Most other targets are changed between subsequent MSA configurations, replaced with new targets to maximize the total object yield. 
 
The CAPERS UDS observations were scheduled between UT 31 December 2024 and 10 January 2025. Two of the 21 planned MSA observations failed due to technical issues with the spacecraft, and 13 more were canceled due to communication restrictions with \jwst\ during the Los Angeles area wildfires in January. Only one UDS pointing was observed using all three planned MSA configurations. Two pointings were observed in two out of three planned MSA configurations, and a final pointing was observed with only one MSA configuration. The remaining three pointings in UDS have not yet been observed. 

Altogether, about 1200 individual targets were observed in these first 8 UDS MSA configurations.
The full details of the target selection, prioritization, and MSA planning will be given in a forthcoming paper. For each MSA configuration, these observations employed a 3-shutter slitlet nod pattern at an aperture position angle (APA) of approximately $201^{\circ}$. Sources were correspondingly assigned 3 shutter slitlets and source-source spectral overlap was avoided whenever possible. Each of the six nods was observed for a single 13 group integration using the \texttt{NRSIRS2} readout pattern.

\subsection{NIRSpec Data Reduction and Calibration}
\label{sec:data_red}

We use the CAPERS team internal data reduction in this work. This reduction largely uses the standard \textit{JWST} pipeline\footnote{\url{https://github.com/spacetelescope/jwst}} \citep{bushouse24} version 1.16.1 and CRDS version \texttt{1312.pmap} with a few modifications, which include the removal of the 1/f noise from the exposures and a modified flat-field file during the \texttt{calwebb\_spec2} stage \citep[see][for more detail]{arrabal_haro23,arrabal_haro23b}.  Observations in three of the UDS pointings incorporated small ($\pm\ 0.05$\arcsec) cross-dispersion dithers between two nod sequences to test potential benefits for outlier rejection and mitigation of artifacts and correlated noise.  These dithers require some additional attention in the pipeline processing. Finally, we run \texttt{calwebb\_spec3} to combine data from the nods and produce our final data products. One-dimensional spectra were extracted for 1196 sources, including both `boxcar' and optimally-weighted extractions for most objects.

Additionally, for sources in slits with compact companions, an alternative nodding pattern was employed, limiting the nodded exposures that are combined in such a way that the shutter containing the contaminant companion is never subtracted from the shutter hosting the primary source. This implies that the background is estimated from fewer shutters in some exposures, but it does not diminish the effective integration time on the source of interest. The result is an asymmetric nodding pattern that avoids oversubtraction in the region of the 2D spectra corresponding to the central source.

\begin{figure*}[h]
\begin{center}
\includegraphics[width=.85\textwidth]{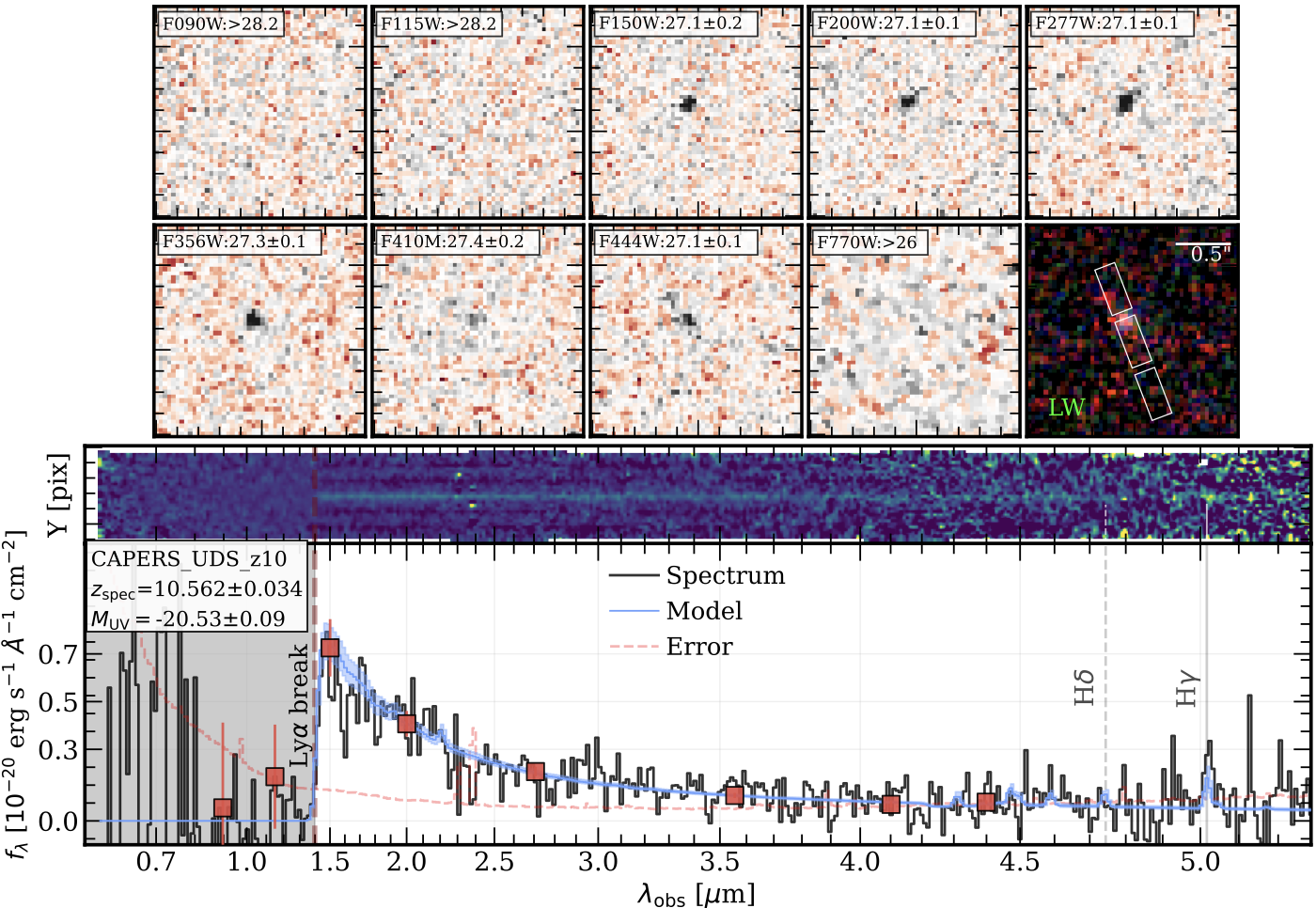}
\includegraphics[width=.85\textwidth]{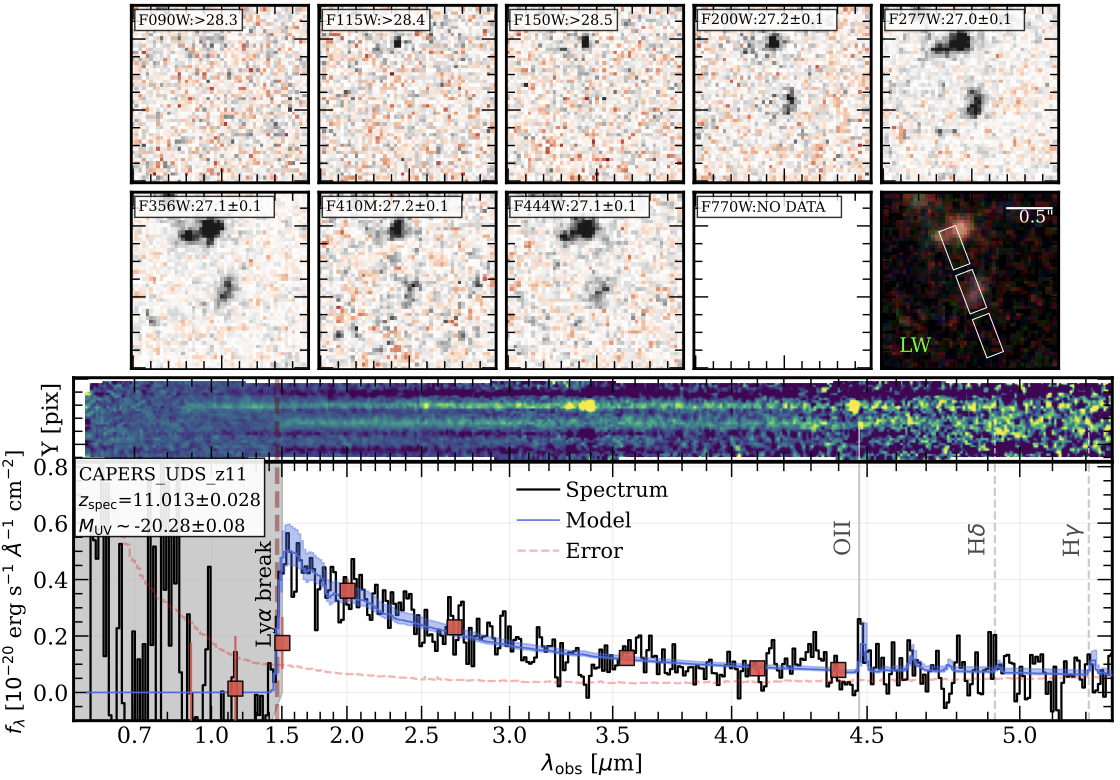}
\caption{\textbf{CAPERS high-$z$ sources.} 
\textbf{Top:} \textit{JWST}/NIRCam and MIRI F770W 2\farcs{0} image stamps and RGB long wavelength color images comprised of the F277W, F356W, and F444W bands, respectively. The MSA slitlet layout (white) is overlaid on the LW image. Each panel shows the total magnitude as presented in the photometric catalog. The spectrum of CAPERS\_UDS\_z11 was processed with an alternative nodding pattern (see end of \autoref{sec:data_red}), to remove the impact of the companion source on the final spectrum.
\textbf{Middle:} 2D MSA PRISM spectrum. 
\textbf{Bottom:} 1D spectra of CAPERS galaxies in the observed frame. We show the spectroscopic data in black, while the 1$\sigma$ uncertainty on the spectrum is in dashed red. Best-fit BAGPIPES model is shown in solid blue, with its uncertainty highlighted by a shaded region. NIRCam photometry is shown with square symbols. Further,  we show the positions and label the prominent emission with significant ($>3\sigma$) detections as solid vertical lines. Emission lines for which we only obtain an upper limit are shown with dashed lines.}
\label{fig:fig_sources}
\end{center}
\end{figure*}

\subsection{Photometry}
\label{sec:phot}

We make use of photometry to select targets for NIRSpec observation, and during our spectroscopic modeling below, both to constrain the overall shape of the spectral energy distribution (SED), and to correct for NIRSpec slit losses.  Our photometric process predominantly follows that described in detail in \citet{finkelstein24}, thus we direct the reader there for more details. For our photometry we use NIRCam and MIRI F770W mosaics provided by the PRIMER team. 
We use Source Extractor \citep{sextractor} to detect sources in a weighted sum of the F277W and F356W images, and to measure photometry in all seven NIRCam bands available from the PRIMER survey (PID 1837, PI: J.Dunlop), as well as MIRI F770W (also from PRIMER), and archival {\it HST}/ACS imaging including the F435W band from PID 16872 (PI: N.Grogin) and the F606W and F814W bands from CANDELS \citep{grogin11,koekemoer11}.  Photometry is measured using small Kron apertures, corrected to total based both on the ratio between the flux in a small Kron and a larger Kron aperture in the F277W image, and a simulation-based residual aperture correction.  Sources are originally identified in a ``cold" mode run, with fainter, smaller sources in a more aggressive ``hot" mode run added in if they were not included in the segmentation map of the cold run \citep[see e.g.][]{rix04}.  Prior to photometry, images with smaller PSFs than the F277W image are convolved with an empirically derived PSF kernel to match the F277W PSF; for larger-PSF images, the photometry is measured without convolution but a correction is derived from the ratio of the flux in the F277W image convolved to the PSF of the larger image to the native F277W image. Photometric redshifts are measured with EAZY \citep{brammer08}, using the provided FSPS templates combined with the updated templates of \citet{larson23}.

\section{Data Analysis}
\label{sec:data_analysis}

\subsection{Spectroscopic Redshifts}
\label{sec:zspec}

The CAPERS team is using a variety of software codes plus interactive inspection and vetting to measure redshifts from the NIRSpec spectra, as will be described in detail in a future paper. 
The present analysis employed an initial review of all 1196 extracted spectra using a modified version of \textsc{msaexp}\footnote{\url{https://github.com/VasilyKokorev/msaexp_OLF}} \citep{msaexp,kokorev24b}. Our procedure first fits a linear combination of \textsc{eazy} \citep{brammer08} models, specifically the \textsc{blue\_sfhz\_13} model subset\footnote{\url{https://github.com/gbrammer/eazy-photoz/tree/master/templates/sfhz}} that contains redshift-dependent SFHs, and dust attenuation values. These models are further complemented by a blue galaxy template, derived from a \textit{JWST} spectrum of a $z=8.50$ galaxy with extreme line equivalent widths \citep{carnall22}. Once we obtain an initial $p(z)$ from the template fitting we perform a second \textsc{msaexp} pass on the sources by fitting sets of Gaussian continuum splines and emission lines templates to our spectrum. As an input parameter, we set \texttt{nsplines=11} and allow to search for the minimal $\chi^2$ value across the prior range set by the template fitting. We leave the position, amplitude and width of emission lines as free parameters, with the latter allowed to vary between 150 -- 800 km/s. The uncertainty is derived via MCMC by resampling the covariance matrix. In these fits we have used the spectra corrected to the photometry discussed in \autoref{sec:phot}. Further, the PRISM dispersion curve is convolved with the input models prior to the fit to account for the wavelength dependent spectral resolution.

Our two-fold approach is especially effective when the number of significantly detected emission lines is limited, like at high redshift, and allows us to constrain the redshift via the break first, and then to refine it and further extract all available information about the emission features in the second pass.

\subsection{Selection of $z>10$ Objects}
\label{sec:highz}
Finally, we select high-$z$ ($z>10$) galaxies from the spectra. Out of 22 $z>10$ candidates placed on the MSA shutters in this first batch of CAPERS-UDS data, we only identify two objects where the high-$z$ nature can be securely confirmed. The analysis of the remaining spectra (also see \autoref{sec:nirspec_setup} regarding failed/canceled observations) is still ongoing, and will be presented in the full CAPERS overview paper once the full data-set is available. (M. Dickinson in prep.).

Initially, CAPERS\_UDS\_z10 (\textsc{ID:136645}) and CAPERS\_UDS\_z11 (\textsc{ID:126973}) were identified at $z_{\rm phot}\simeq9.97$ and $z_{\rm phot}\simeq11.95$ via the \textsc{EAZY} photometric fitting. The two objects were observed between January 1st and 9th 2025. In \autoref{fig:fig_sources} we show the shutter positions overlaid on  2\farcs{0} cutouts of the high-$z$ sources. To mitigate the contamination from a low-$z$ source in one of the CAPERS\_UDS\_z11 slitlets, we only combine the nodded exposures in a way that the contaminant is never subtracted from the spectrum of the primary source, as described in \autoref{sec:data_red}. Both of the objects presented in this work were each observed in two MSA configurations for a total of 11379\,s each. Further, we do not observe strong emission lines detected at high significance in either object, however a clear Ly$\alpha$ break present in both sources allows us to securely derive a redshift of $z_{\rm spec}=10.56\pm0.04$ for  CAPERS\_UDS\_z10 and $z_{\rm spec}=11.00\pm0.04$ for  CAPERS\_UDS\_z11.

\subsection{Emission Lines}
\label{sec:lines}
Both spectra exhibit prominent Ly$\alpha$ breaks at $\lambda_{\rm rest}\simeq1216$ \AA, which are clearly visible in both the 1D and 2D spectra (\autoref{fig:fig_sources}) with a high level of significance. There is no indication of Ly$\alpha$ emission, as has been reported in a handful of other $z>10$ galaxies \citep{bunker23,curtislake23,witstok2025}, though the resolution of the prism prohibits detection of faint emission lines at the blue end of the spectrum. To further investigate the spectral properties of these galaxies, we conducted a detailed inspection of both the 1D and 2D PRISM spectra, searching for prominent emission lines above the noise level.
To do that we use the 1D Gaussian \textsc{msaexp} fits to our spectra. Despite a $\sim3\sigma$ detection of the continuum in both objects across the majority of the spectrum, we identify only two emission lines: a significant ($>4 \sigma$) detection of H$\gamma$ in CAPERS\_UDS\_z10, and a $\sim5\sigma$ \oii\ $-$ 3727 \AA\, line in CAPERS\_UDS\_z11, From both lines we are thus able to measure more precise spectroscopic redshifts of $z_{\rm spec}=10.562\pm0.034$ and $z_{\rm spec}=11.013\pm0.028$, respectively, fully consistent with our previous measurement from the Ly$\alpha$ break. We will adopt line redshifts for both sources throughout our work. With \textsc{msaexp} we measure the rest-frame 
equivalent width (EW$_0$) of the detected H$\gamma$ in CAPERS\_UDS\_z10 at $74\pm18$ \AA.

The second object,  CAPERS\_UDS\_z11, appears devoid of any significantly detected Balmer lines.
The absence of strong spectral features is not unexpected at high redshifts, as the strong emission lines such as the [\oiii] doublet and H$\beta$ shift our of the NIRSpec wavelength coverage ($\gtrsim5.3$ $\mu$m). This leaves us only with higher order (and fainter) Balmer lines and typically-weak metal lines in the observed spectrum range. For instance, the current record-breaking high-$z$ galaxies presented in \citet{carniani24} also exhibit a lack of prominent emission lines in the rest-optical for that reason. In addition, the expected position of the H$\gamma$ line at $z\sim11$ would fall into a much noisier part of the spectrum, further complicating its detection. We, however, do observe an \oii\ line at EW$_{0}=47\pm10$ \AA\, in this object, although given the shape of this line and its proximity to a trough in the continuum the true SN of this feature is likely below 5. Further, we also derive upper limits on the flux and EW$_0$ of both H$\delta$ and H$\gamma$ for  CAPERS\_UDS\_z11. 
We report the detected emission line fluxes, EW$_0$ and upper limits in \autoref{tab:tab2}.

\begin{deluxetable}{cCC}[]
\tabcolsep=2mm
\tablecaption{\label{tab:tab1} CAPERS UDS high-$z$ sample. }.
\tablehead{Parameter &  \mathrm{CAPERS\_UDS\_z10} &  \mathrm{CAPERS\_UDS\_z11}}
\startdata
RA [deg] & 34.456023 &  34.264439\\
Dec [deg] & -5.121952 & -5.096232 \\
$z_{\rm spec}$ &  $10.562\pm0.034$ & $11.013\pm0.028$ \\
$M_{\rm UV}$ [AB mag] & $-20.53\pm0.09$ & $-20.28\pm0.08$ \\
$R_{\rm eff}$ [pc] & $420\pm70$ & $560\pm61$\\
\hline
\multicolumn{3}{c}{\textsc{BAGPIPES Fit}} \\
\hline \hline
$\beta$ & $-2.27\pm0.12$ & $-1.71\pm0.12$ \\
SFR$_{10}$ [M$_\odot$/yr] & $18\pm5$ &  $43\pm10$ \\
SFR$_{100}$ [M$_\odot$/yr] & $9\pm5$ & $22\pm5$ \\
log$_{10}$(SFR$_{10}$/SFR$_{100}$) & $0.3\pm0.2$ & $0.3\pm0.5$\\
log$_{10}$($M_*/M_\odot$) & $8.3\pm0.2$ & $8.7\pm0.2$ \\
$A_{\rm V}$ & $0.31\pm0.10$ & $0.74\pm0.11$ \\
Age [Myr] & $<18$ & $<20$ \\
\hline
\multicolumn{3}{c}{\textsc{Empirical}} \\
\hline \hline
SFR$_{10}$ [M$_\odot$/yr] & $43\pm10$ & $<80$ \\
SFR$_{100}$ [M$_\odot$/yr] & $20\pm3$ & $43\pm5$ \\
log$_{10}$(SFR$_{10}$/SFR$_{100}$) & $0.3\pm0.1$ & $<0.3$ \\
\enddata
\begin{tablenotes}
\end{tablenotes}
\end{deluxetable}

\subsection{SED Fitting}
\label{sec:bagpipes}
So far we have used \textsc{msaexp} to measure spectroscopic redshifts as well as the intensities of the few spectral lines we have detected. 
We have noted that the strong H$\gamma$ line in CAPERS\_UDS\_z10 could indicate that the galaxy light is dominated by a burst of star formation \citep[e.g. see][]{leitherer14}. To test this hypothesis we perform stellar population modeling with the SED fitting code \textsc{BAGPIPES} \citep{carnall18,carnall19}, which simultaneously models both the photometry and full spectrum to constrain a variety of physical properties, including the star-formation history.

We adopt a non-parametric star-formation history (SFH) with the `bursty continuity' prior from \citet{tacchella22}, using eight time bins where the SFR is fit to a constant value in each bin. The first four bins are set to lookback times of 0-3 Myr, 3-10 Myr, 10-30 Myr, and 30-100 Myr, while the last four bins are logarithmically spaced between 100 Myr and $t_\mathrm{max} = t_\mathrm{universe}(z=z_\mathrm{spec}) - t_\mathrm{universe}(z=20)$. We fixed the redshift at the spectroscopic redshift to remove any photometric uncertainties that may propagate into the SED fitting. Dust attenuation is assumed to follow the \citet{calzetti00} law and the rest-frame V-band attenuation is fit (log-uniform prior) in the range $A_{\rm V} = 0.001 - 3$ mag. Stellar nebular emission is included with an ionization parameter in the range $-4 \leq$ log $U \leq 0$. Our log $U$ upper limit was chosen to accommodate galaxies with high ionization values, as is common at high redshift \citep[e.g. see][]{carnall22}. We allow for an intrinsic stellar velocity dispersion in the range 50--500 km/s, though we provide the BAGPIPES fits with the NIRSpec PRISM dispersion curve to convolve the resolution of the BAGPIPES models to the resolution of the input spectral data. The metallicity is allowed to vary between 0.001--1 $Z_\odot$ (log-uniform prior for both). To account for differences between the spectra and photometry due to slit losses, we incorporate parameters into the BAGPIPES fitting that handle the modeling of noise and the calibration to the photometry using the prescription outlined in \citet{carnall19}. 

\begin{figure*}
\begin{center}
\includegraphics[width=.86\textwidth]{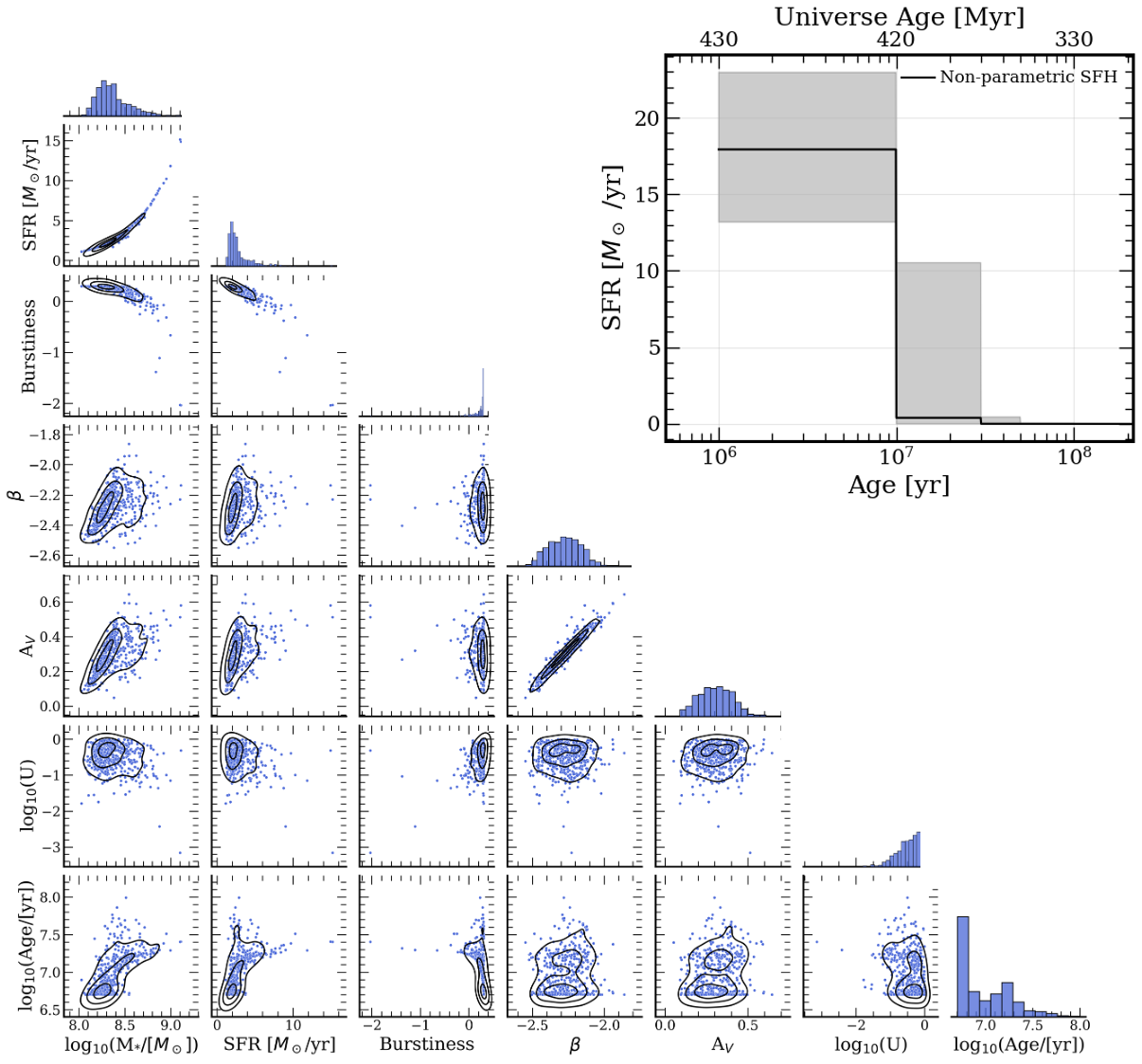}
\caption{\textbf{BAGPIPES Posteriors and SFH for  CAPERS\_UDS\_z10.} \textbf{Left:} Posterior probability distributions obtained with \textsc{BAGPIPES} for the non-parametric SFH fit. The columns (left to right) show the 
$M_*$, SFR$_{100}$, Burstiness (defined as log$_{10}$SFR$_{10}$/SFR$_{100}$), UV slope $\beta$, dust extinction $A_{\rm V}$ , ionization parameter log$_{10}U$ and the mass weighted age. We plot the 1D posterior distributions of each parameter along the diagonal. On 2D scatter plot we show the joint posterior distributions with contours representing the 1, 2 and $3\sigma$ intervals. \textbf{Right:} The best-fit non-parametric SFH. The shaded region shows the $\pm1\sigma$ uncertainty on the fit.}
\label{fig:fig_corner1}
\end{center}
\end{figure*}

\begin{figure*}
\begin{center}
\includegraphics[width=.86\textwidth]{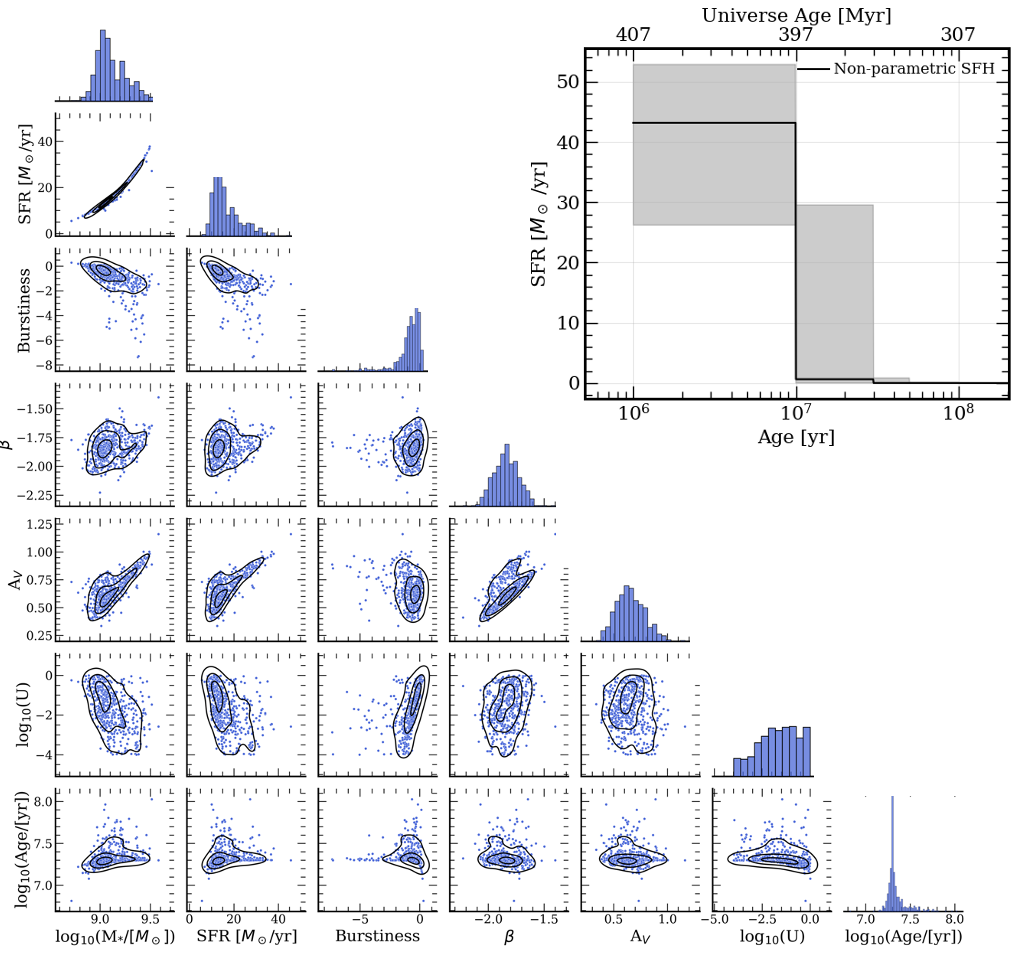}
\caption{\textbf{BAGPIPES Posteriors and SFH for  CAPERS\_UDS\_z11.} Same as \autoref{fig:fig_corner1}.}
\label{fig:fig_corner2}
\end{center}
\end{figure*}

To test other star formation history models we repeat the same procedure but this time restrict the burstiness parameter in the non-parametric SFH model to be strictly positive, replicating a rising SFH. We also tested a declining SFH by adopting an exponential SFH model. The parameter bounds for the dust and nebular priors are exactly the same as the non-parametric SFH. The bounds for the exponential SFH parameters are: a uniform prior of age to be 0.001--1 Gyr, uniform prior in $\tau$ to be 0.002--11, a log$_{10}$ prior for metallicity ranging between 0.001--1 $Z_\odot$. The posteriors and SFHs for our galaxies are presented in \autoref{fig:fig_corner1} and \autoref{fig:fig_corner2}.

From the \textsc{BAGPIPES} modeling, we find that  CAPERS\_UDS\_z10 is well characterized by a rising non-parametric SFH ($\Delta \chi^2>3.0$ compared to declining SFH model), indicative of a current starburst. This conclusion is supported by the ratio of SFRs measured over the recent 10 Myr (SFR$_{10}$) and 100 Myr (SFR$_{100}$) timescales, which yields log$_{10}$(SFR$_{10}$/SFR$_{100}$) $\sim 0.3$ (which is consistent with the value derived directly from the spectrum, as will be discussed in \autoref{sec:beta_slope}). The object also exhibits a relatively blue UV spectral slope ($\beta \sim -2.27$) and a modest level of dust obscuration ($A_{\rm V} \sim 0.3$~mag).

In contrast, results for the second object, CAPERS\_UDS\_z11, are less definitive. The non-parametric model prefers a recent burst, similarly to the first galaxy, however the uncertainty on the fit is larger due to a lack of securely detected Balmer lines. To test the validity of this fit, we check the alternative SFH parameterizations fit to this object and find that the data are non-parametric SFH is marginally ($\Delta \chi^2\sim1-2$) preferred, compared to the rising and exponentially declining star formation histories. This ambiguity arises primarily due to the lack of Balmer series lines, limiting our ability to constrain the recent star formation history reliably. We reflect this in our uncertainty on the measurement where we find log$_{10}$(SFR$_{10}$/SFR$_{100})=0.3\pm0.5$. With the current data, it remains uncertain whether  CAPERS\_UDS\_z11 is undergoing a recent burst, or is ``napping'' in a period of temporary quiescence following a burst \citep[e.g.][]{looser24, cole25}.

Additionally,  CAPERS\_UDS\_z11 is notably redder ($\beta \sim -1.7$) compared to CAPERS\_UDS\_z10 with a \textsc{BAGPIPES} fit implying a higher dust attenuation ($A_{\rm V} \sim 0.7$~mag). These properties are consistent across all SFH scenarios we explore. A summary of all derived physical parameters from our \textsc{BAGPIPES} analysis is provided in \autoref{tab:tab1}.

\begin{figure*}
\begin{center}
\includegraphics[width=.75\textwidth]{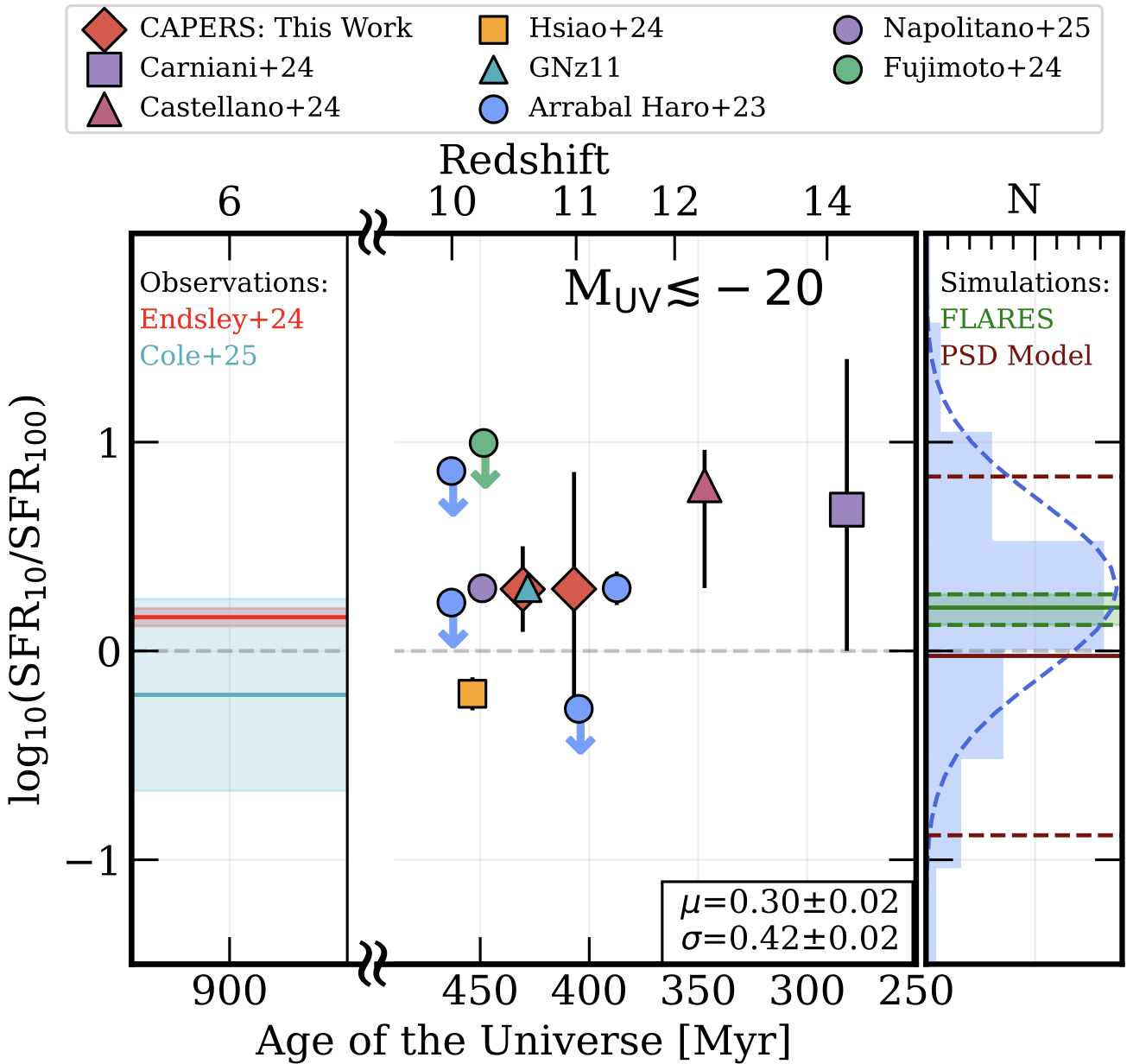}
\caption{\textbf{Bursting Star Formation of bright of $z>10$ galaxies.} The ratio between SFR$_{10}$ and SFR$_{100}$ as a function of the age of the Universe. Alongside the CAPERS (red) galaxies, we show all the literature results for $M_{\rm UV}<-20$ high-$z$ galaxies which report on the recent SFR \citep{carniani24,hsiao23,hsiao24,bunker23,alvarezmarquez24,fujimoto24_uncover,napolitano25}. The CEERS high-$z$ galaxies presented in \citep{arrabal_haro23,arrabal_haro23b}, and bright $z\sim10$ galaxies presented in \citet{napolitano25} and \citet{fujimoto24_uncover} do not report SFR$_{10}$, and were thus fit with the same \textsc{BAGPIPES} set-up as discussed in \autoref{sec:bagpipes}. In the right panel we show burstiness distribution created by MCMC resampling all the available data. The dashed blue line shows the best-fit Gaussian to the burstiness distribution, with the best-fit mean ($\mu$) and standard deviation ($\sigma$) listed in the inset panel. We additionally show results from the PSD toy model (maroon) and hydrodynamical simulations \citep[FLARES;][in green]{lovell21,wilkins23_v,wilkins23_vii} of burstiness at high-$z$. Finally, in a subpanel on the left we show medians (solid lines) and 68 \% confidence intervals (shaded regions) of observational constraints on burstiness of bright $z\sim6$ galaxies from \citet{endsley24} and \citet{cole25}.}
\label{fig:burst}
\end{center}
\end{figure*}

\subsection{Empirical UV Properties}
\label{sec:beta_slope}

We estimate the absolute UV magnitude for each galaxy from the photometry-calibrated spectrum at $\lambda_{\rm rest}\sim1500$ \AA, and find that our targets have $M_{\rm UV}=-20.53\pm0.09$ for  CAPERS\_UDS\_z10 and $M_{\rm UV}=-20.35\pm0.08$ for  CAPERS\_UDS\_z11. On top of deriving the UV-slope $\beta$ with \textsc{BAGPIPES}, we also compute it by assuming $f_{\lambda}\sim \lambda^{\beta}$ and deriving it directly from the spectrum within the $\lambda_{\rm rest}=1250 – 3000$ \AA\, range. From the photometry corrected spectra we directly measure $\beta=-2.26\pm0.11$ and $\beta=-1.90\pm0.12$ for CAPERS\_UDS\_z10 and CAPERS\_UDS\_z11, respectively, fully consistent with the $\beta$ values already derived with \textsc{BAGPIPES}, within $<1\sigma$. 

As a check on the derived SFR$_{10}$/SFR$_{100}$, we perform an independent empirical estimation based on emission line fluxes (or their upper limits) and the UV magnitude ($M_{\rm UV}$). We assume Case B recombination conditions, adopting intrinsic line ratios of (H$\alpha$/H$\gamma)_{\rm int}\sim6.1$ and (H$\alpha$/H$\delta)_{\rm int}\sim10.6$ \citep{osterbrock89}. Using the dust attenuation derived from \textsc{BAGPIPES}, we correct the observed line fluxes (or upper limits) to estimate the intrinsic H$\alpha$ luminosity, and convert the observed $M_{\rm UV}$ to the intrinsic one. So far, there are only 3 direct H$\alpha$ detections at $z>10$, so we can attempt to directly compare our Case B converted luminosities to them. We find that our dust-corrected H$\alpha$ luminosities are $L_{\rm H\alpha}\sim6\times10^{42}$ erg/s for CAPERS\_UDS\_z10 
and an upper limit $L_{\rm H\alpha}\lesssim6\times10^{42}$ erg/s for CAPERS\_UDS\_z11. Interestingly, comparing the $L_{\rm H\alpha}$ to the existing high-$z$ samples, we find a value that is $\times5$ higher than that in MACS0647-JD \citep{hsiao24}, and is much more comparable to GHZ2 \citep{zavala25} and GNz11 \citep{alvarezmarquez24}.
We will expand on it futher in \autoref{sec:burstiness}. Finally, our intrinsic H$\alpha$ luminosities and the $M_{\rm UV}$ are then converted to SFR$_{10}$ and SFR$_{100}$, respectively, following the relations provided by \citet{Kennicutt98}.

From the detected H$\gamma$ emission line in  CAPERS\_UDS\_z10, we empirically determine log$_{10}$(SFR$_{10}$/SFR$_{100}$) $= 0.3\pm0.1$, fully consistent with the \textsc{BAGPIPES} result. For  CAPERS\_UDS\_z11, neither H$\gamma$ nor H$\delta$ lines are detected, limiting our estimate to an upper bound on SFR$_{10}$. Given that the expected H$\gamma$ line position lies too close to the upper observed wavelength limit of the PRISM, we adopt the H$\delta$ flux upper limit (\autoref{tab:tab2}) to constrain the SFR$_{10}$. This yields log$_{10}$(SFR$_{10}$/SFR$_{100}$) $< 0.3$, which is consistent with, albeit lower than the limit derived by \textsc{BAGPIPES}.

\subsection{Size Measurements}
\label{sec:size}
Cutouts presented in \autoref{fig:fig_sources} show that while our sources are compact they are still clearly resolved.
We measured the effective radius ($R_{\rm eff}$) to compare the sizes of CAPERS $z>10$ galaxies to other galaxies at similarly early epochs, modeling the galaxies with \textsc{GALFIT} \citep{peng02,peng10}. When fitting we take into account the effects of the PSF, which we have measured empirically from the bright stars in the field. The same PSFs were also used when constructing the photometric catalog described in \autoref{sec:phot}. We model the object with a S\'ersic \citep{sersic} profile where the source position, brightness, effective radius, S\'ersic index, and axis ratio are allowed to vary. 

We perform this procedure on our highest SNR band, F277W, to ensure optimal S/N per pixel is achieved to accommodate robust size measurements, a procedure often employed for other $z>10$ galaxies \citep[e.g. see][]{arrabal_haro23,kokorev24_glimpse}. Moreover, while the F277W band has a $\times2$ poorer resolution compared to the F200W, the higher S/N (per pixel) in the LW band means that extended structure can be detected and modeled in a more robust way.
We find that the on-the-sky sizes of our sources are $R_{\rm eff}=0.10\pm0.02$\arcsec\, for  CAPERS\_UDS\_z10 and $R_{\rm eff}=0.14\pm0.02$\arcsec\, for CAPERS\_UDS\_z11. Both of our sources appear to be resolved, with 
CAPERS\_UDS\_z11 in particular showing potential signs of interactions/merger. Taking the redshift into account, we convert our angular sizes to physical effective radii. We obtain $R_{\rm eff} = 420\pm70$ pc for  CAPERS\_UDS\_z10 and $R_{\rm eff} = 460\pm61$ pc  CAPERS\_UDS\_z11. These sizes are also listed in \autoref{tab:tab1}.

\section{Results and Discussion} 
\label{sec:res}
\subsection{Stellar Population Properties}
While only observed with modest ($\sim3.2$ hours) integration times, these targets are sufficiently bright to be detected at $>3\sigma$ level across all spectral pixels.
While this, plus the detection of certain emission line features, is perfectly adequate to constrain the redshift,  properties of the most recent episode of star formation, such as UV-slope, age and dust obscuration, deriving any further physical parameters such as stellar mass or metallicity would not be reasonable due to the lack of rest-optical continuum and a sufficient number of detected emission lines. As such, we will only focus on stellar population parameters that can be derived from the rest-UV alone.

We find that both of our objects are bright, with a median $M_{\rm UV}\sim-20.4$, therefore confirming these two CAPERS galaxies at $z=10.562$ and $z=11.013$, which raises the number of UV-bright ($M_{\rm UV}\sim-20$) galaxies at $z>10$ to a total of 12 \citep{arrabal_haro23,arrabal_haro23b,castellano24,carniani24,hsiao23,hsiao24,oesch16,bunker23,maiolino23,napolitano25,fujimoto24_uncover}. We now would like to explore what other similarities, apart from brightness and their high-$z$ nature, and any significant differences these objects have.

It does appear that while dust attenuation, as derived by \textsc{BAGPIPES}, is low, it is not negligible.  For CAPERS\_UDS\_z10 we find  $\beta\simeq-2.27$ and $A_{\rm V}\simeq0.3$ which exactly match the ones observed in JADES-GS-z14-0 \citep{carniani24}, assuming that the \citet{calzetti00} attenuation law is valid at high redshift \citep[used in other high-$z$ galaxy works e.g.][]{carniani24}. Further, the BAGPIPES fit includes nebular continuum so it is unlikely that the reddening we derive is being artificially induced. While we report no UV lines in CAPERS\_UDS\_z10, \citet{carniani24} find a tentative ${\rm {C}\,\textsc{iii]}}$ line for JADES-GS-z14-0.  Given that our object is $\sim0.3$ mags fainter, it is likely that the S/N of our spectrum is not sufficiently high. On the other hand,  CAPERS\_UDS\_z11 has a notably shallower UV slope - $\beta\simeq-1.71$ and consequently a larger $A_{\rm V}\sim0.74$. If true, this would make it one of the dustiest high-$z$ galaxies observed by \textit{JWST}, and could in part explain the lack of emission lines. It is worth noting that we still lack data beyond $\lambda_{\rm rest}\sim 4000$ \AA, and therefore constraining dust attenuation solely relies on the UV slope. Moreover, a lack of detection of typical metallicity-constraining lines further complicates a robust estimate of $A_{\rm V}$. Further observations of the dust continuum in the CAPERS (and other) high-$z$ galaxies with e.g. ALMA would be needed to more accurately constrain their dust content.

Finally, it is also worth noting that while both of the objects presented in this work are UV luminous, their significant level of dust attenuation shows that bright galaxies at $z\gtrsim10$ can be somewhat attenuated \citep[c.f. see][]{mitsuhashi25}, casting doubt on feedback-based dust removal as the only cause of the high bright end of the UVLF \citep[e.g.][]{ferrara23}.

\subsection{Compact and Extended Galaxies at $z>10$}
\label{sec:compactness}
We further compare our measured effective radii with empirically derived sizes of spectroscopically confirmed galaxies at $z > 10$, as well as predictions derived from extrapolating scaling relations established at lower redshift. The two galaxies examined in our analysis exhibit extended morphologies, characterized by effective radii of approximately $R_{\rm eff} \sim 500$ pc. In addition CAPERS\_UDS\_z11 appears to have multiple components and faint extended features, likely hinting at its interacting/merging nature (and more tentatively also CAPERS\_UDS\_z10). These measurements align closely with those observed in the similarly luminous ($M_{\rm UV} < -20$) high-redshift galaxy sample from CEERS, as recently presented by \citet{arrabal_haro23,arrabal_haro23b}. Furthermore, the sizes of CAPERS galaxies are consistent within roughly $1\sigma$ uncertainty with the current record holder for highest confirmed spectroscopic redshift—JADES-GS-z14-0 \citep{carniani24}.

Interestingly, extended morphology does not universally characterize luminous galaxies at these early epochs. Objects such as GNz11 \citep{oesch16,maiolino23,bunker23} and GHZ2 \citep{castellano24,ono23,morishita22} are significantly more compact, with radii constrained to $R_{\rm eff} \lesssim 100$ pc. This striking contrast in morphologies among similarly bright, early-universe galaxies raises important questions: Does the morphological diversity observed at these extreme redshifts provide insights into the underlying mechanisms driving their remarkable UV luminosities? Can the differences in galaxy sizes and structures illuminate pathways governing their early evolutionary histories?

While a statistical census of ionization properties and sizes at high redshift remains limited by data volume, emerging evidence points to possible physical drivers behind observed trends. The presence of high-ionization UV lines in compact objects may be driven by AGN activity or extremely compact star formation \citep{alvarezmarquez24,castellano24,zavala25}. In contrast, the enhanced brightness of extended galaxies at $z>10$ may result from high star formation efficiency and/or stochastic star formation potentially triggered by mergers—similar to what is observed in $z\sim7$ systems \citep{harikane24}.

Theory predicts higher merger rates in the early Universe \citep[e.g.][]{fakhouri10,rodriguezgomez15} compared to those at $z\sim6-7$. Moreover, major merger timescales at high redshift are shorter than the age of the Universe, which, as \citet{harikane24} argue, may imply that most $M_{\rm UV}$-bright galaxies have undergone at least one such event.

The galaxies presented in this study exhibit extended morphologies and notably distinct spectral features compared to compact high-$z$ objects, which typically display stronger UV emission lines. We specifically would like to note signs of potential interactions/mergers present in both objects, particularly in CAPERS\_UDS\_z11 which exhibits two distinct clumps in \autoref{fig:fig_sources}. Currently however, fully characterizing the relationship between galaxy sizes and UV line intensities at $z>10$ remains challenging due to limited sample sizes. Addressing this gap is one of the primary objectives of the CAPERS survey, which is anticipated to deliver tens of additional high-$z$ galaxy spectra upon completion, significantly advancing our understanding of how early galaxy properties affect their morphology, and vice-versa.

\begin{deluxetable}{cccc}[]
\tabcolsep=3mm
\tablecaption{\label{tab:tab2} Emission Line Properties.}.
\tablehead{ID & \oii &  H$\delta$ & H$\gamma$  \\
 & 3727 \AA & 4101.7 \AA & 4340.4 \AA }
\startdata
\multicolumn{4}{c}{Line Flux [10$^{-20}$ erg s$^{-1}$ cm$^{-2}$]} \\
\hline
UDS\_z10 & -- & -- & $54.6\pm14.0$ \\
UDS\_z11 & $37.2\pm7.3$ & $<24$ & $<20$ \\
\hline
\multicolumn{4}{c}{Rest Frame EW [\AA]} \\
\hline \hline
UDS\_z10 & -- & -- & $74\pm18$ \\
UDS\_z11 & $46\pm10$& $<19$ & $<10$ \\
\enddata
\begin{tablenotes}
\end{tablenotes}
\end{deluxetable}

\subsection{Bursting Star Formation}
\label{sec:burstiness}
We will now explore how prevalent bursting star formation is at the bright end ($M_{\rm UV}<-20$) of the high-$z$ UVLF. To achieve this, we complement our derived burstiness metric, log$_{10}$(SFR$_{10}$/SFR$_{100}$), with similarly luminous galaxies reported in the literature showing evidence of bursting star formation \citep{alvarezmarquez24,bunker23,carniani24,castellano24,hsiao23,hsiao24,zavala25}, as well as the bright $z>10$ CEERS galaxies from \citet{arrabal_haro23,arrabal_haro23b}. Since the SFR$_{10}$ values are not reported for the CEERS objects, we have re-fit them using \textsc{BAGPIPES} following our methodology outlined in \autoref{sec:bagpipes}. In cases where burstiness could not be robustly quantified, only upper limits are provided. We classify a galaxy as ``bursting'' if log$_{10}$(SFR$_{10}$/SFR$_{100})>0$ \citep[e.g., see][]{endlsey24}; our results are shown in \autoref{fig:burst}.

To statistically assess the significance of stochastic star formation among these 12 galaxies, we performed a Monte Carlo Markov Chain (MCMC) resampling of each derived log$_{10}$(SFR$_{10}$/SFR$_{100})$ value within its uncertainty, repeating this process $\sim10^8$ times. The resulting distribution is shown in the right panel of \autoref{fig:burst}. A simple Gaussian fit to this distribution shows that indeed the sample of these galaxies can be considered bursting with a mean of $\mu=0.30\pm0.02$~dex, albeit with a considerable scatter $\sigma=0.42\pm0.02$~dex. 

Before discussing the implications of our measurement, let us contrast our findings against simulated predictions of burstiness in bright galaxies at high redshift. We will start with analytic predictions. The median and width of the burstiness distribution can be interpreted in terms of an analytic model, where fluctuations on $\log$ SFR$(t)$ are normally distributed with a known  power-spectrum distribution (PSD,~\citealt{caplar19,iyer20}). The log$_{10} \rm \left(SFR_{10}/SFR_{100}\right)$ statistic corresponds to a weighted integral over the PSD, and thus can be used to constrain the amplitude and timescale of burstiness. Following the methods described in \citet{sun_g24}, we focus on the bright end of the UVLF at $z>10$ and derive a median log$_{10} \rm \left(SFR_{10}/SFR_{100}\right)\sim-0.005$ (in agreement with zero in this toy case) and a scatter of $0.8$ for a PSD amplitude $\sigma=4$ and timescale $\tau_{\rm decor} = 5$ Myr, which should also give rise to a high UV variability, though these parameters are degenerate. We plot the median and 68\% confidence intervals of this distribution in \autoref{fig:burst}. We complement this comparison by including the burstiness metric results from the hydrodynamical FLARES simulations \citep{lovell21,wilkins23_v,wilkins23_vii}, focusing on the bright end of the UVLF ($M_{\rm UV}\lesssim -20$) at $z\sim11$, where the majority of our samples lies. In contrast to the simplified PSD toy model, the FLARES simulations strongly suggest that 
bright galaxies at high redshift are indeed bursting, with a tight distribution characterized by log$_{10}$(SFR$_{10}$/SFR$_{100}$) $\sim 0.2$ and $\sigma \sim 0.08$. Our observed median burstiness is broadly consistent with both simulation predictions within $1\sigma$. However, the large scatter in our measurements—driven primarily by the limited number of spectroscopically confirmed galaxies in our sample—makes it difficult to assess the true statistical significance of this agreement.

Nonetheless, our results suggest that, on average, bright galaxies at $z > 10$ are in a bursting phase, with SFR$_{10}$ exceeding SFR$_{100}$ by approximately a factor of two. Although the NIRSpec/MSA selection function is challenging to fully characterize, spectroscopically confirmed samples of bright galaxies are expected to be largely complete. Moreover, these galaxies are selected based on their bright $M_{\rm UV}$, rather than emission line intensities, implying that the selection is not intrinsically biased toward bursting systems—or against them.

To evaluate whether stochastic star formation may contribute to the observed overabundance of UV-bright galaxies at these redshifts, it is useful to compare the prevalence of burstiness at $z > 10$ to that at lower redshifts. We begin with the work by \citet{endlsey24}, who used a photometrically selected sample of H$\alpha$ emitters at $z \sim 6$ and found a very tight burstiness distribution of log$_{10}$(SFR$_{3}$/SFR$_{50}$) $\sim 0.16 \pm 0.04$ for galaxies with $M_{\rm UV} < -20$. In a complementary study, \citet{cole25} analyzed a smaller, spectroscopically confirmed sample of bursting galaxies at similar redshifts, finding a broader distribution centered at $\sim -0.21 \pm 0.46$, with the large scatter driven by low number statistics—similar to our case.

Both observational results at $z \sim 6$ are $\sim 0.1$ dex and $\sim 0.5$ dex lower, respectively, than the median burstiness derived in our work. This difference may hint that, at least among $M_{\rm UV} < -20$ galaxies, burstiness is a more prominent feature at $z > 10$ than at later times, suggesting that stochastic star-formation could at least in-part drive the overabundance of UV-bright objects at early times.
However, given the large scatter in our measured distribution, it remains difficult to determine with high confidence whether this trend is significant. To reach the level of precision afforded by both photometric samples and simulations, the number of confirmed high-redshift galaxies would need to increase by at least a factor of $\sim9$ (i.e., $1/\sqrt{N}$), in order to reduce the scatter in our observed distribution by a factor of $\sim3$, to match lower uncertainties found in e.g. \citet{endsley24}. Continued spectroscopic follow-up of bright galaxies at $z>10$ with \textit{JWST} will be critical in establishing whether bursting star formation is indeed a defining feature of the earliest galaxies.

\section{Conclusions} \label{sec:concl}
We present spectroscopic identification of two $z>10$ galaxies obtained with the first 14\% of the CAPERS MSA observations.
Both objects present a clear Ly$\alpha$ break which results in unambiguous redshifts of $z=10.562$ and $z=11.013$. We examine physical properties of both galaxies via SED fitting and empirically as well as morphologies and star formation histories, finding evidence for recent bursting star formation in CAPERS\_UDS\_z10, thanks to the detection of strong (EW$_0\sim70$ \AA) H$\gamma$ emission. The latter object CAPERS\_UDS\_z11, lacks any Balmer emission lines detected at high significance, preventing us from securely confirming or ruling out an ongoing burst, however it does present a tentatively detected \oii\ emission line.

We combine our sample with all other bright $M_{\rm UV}<-20$ galaxies at $z>10$ from recent literature \citep{arrabal_haro23,arrabal_haro23b,alvarezmarquez24,bunker23,carniani24,castellano24,hsiao23,fujimoto24_uncover,napolitano25} and explore what role 
stochastic star formation has had in shaping early galaxy formation. We find that the bright end of the UVLF at $z>10$ is likely to be dominated by galaxies observed during an ongoing starburst, as  characterized by a median log$_{10}$(SFR$_{10}$/SFR$_{100})\sim0.30$~dex, with a scatter of $\sigma\sim0.42$~dex.

Determining the “burstiness” of a galaxy typically requires a well-constrained SFH to reliably estimate the SFR within the last 10 Myr. However, this becomes particularly challenging at $z>10$, as we have discussed, where typical strong emission lines such as H$\beta$ and [\oiii], redshift beyond NIRSpec spectral coverage, thus making such measurements increasingly uncertain. While one can still place upper limits on burstiness either through SED fitting or empirically via non-detections of emission lines, these approaches are inherently limited and subject to significant degeneracies—underscoring the need for deeper (and wider) spectroscopic data to robustly trace recent star formation activity at early cosmic times. In the next few months CAPERS is expected  to double the number of galaxies at $z>10$, thus yielding us the much needed statistical certainty to determine the true impact (or lack thereof) of bursting SF on the early galaxy physics.

\acknowledgements
The authors would like to thank Richard Ellis for insightful discussions about bursting star-formation at high redshift,  the PRIMER team for carrying out \jwst\ imaging of the UDS, and notably Dan Magee for reducing the PRIMER NIRCam data. VK acknowledges support from the University of Texas at Austin Cosmic Frontier Center.  AT and SF acknowledge support from NASA through grant JWST-GO-6368. LN acknowledges support from grant ``Progetti per Avvio alla Ricerca - Tipo 1, Unveiling Cosmic Dawn: Galaxy Evolution with CAPERS" (AR1241906F947685). PGP-G acknowledges support from grant PID2022-139567NB-I00 funded by Spanish Ministerio de Ciencia e Innovaci\'on MCIN/AEI/10.13039/501100011033, FEDER {\it Una manera de hacer Europa}. This work is based on observations made with the NASA/ESA/CSA \textit{James Webb Space Telescope}, obtained at the Space Telescope Science Institute, which is operated by the Association of Universities for Research in Astronomy, Incorporated, under NASA contract NAS5-03127. Support for program number GO-6368 was provided through a grant from the STScI under NASA contract NAS5-03127. The data presented in this article were obtained from the Mikulski Archive for Space Telescopes (MAST) at the Space Telescope Science Institute. These observations are associated with program \#6368. 

\software{BAGPIPES \citep{carnall19}, BEAGLE \citep{chevallard16}, EAZY \citep{brammer08}, GALFIT \citep{peng02}, grizli \citep{grizli}, sep \citep{sep}, SExtractor \citep{sextractor}}

\facilities{\jwst, \hst}

\clearpage

\bibliographystyle{aasjournal}
\bibliography{refs}

\end{document}